\documentclass[prl,twocolumn,showpacs]{revtex4}
\usepackage{graphicx}
\usepackage{times}
\usepackage{bm}

\def\be{\begin{equation}}
\def\ee{\end{equation}}
\def\bea{\begin{eqnarray}}
\def\eea{\end{eqnarray}}
\def\bse{\begin{subequations}}
\def\ese{\end{subequations}}

\def\be{\begin{eqnarray}}
\def\ee{\end{eqnarray}}

\begin{document}

\title{Time-reversal symmetry breaking by \boldmath{${(d+id)}$}
density-wave state in underdoped cuprate superconductors}
\author{Sumanta Tewari}
\author{Chuanwei Zhang}
\author{Victor M. Yakovenko}
\author{S. Das Sarma}
\affiliation{Condensed Matter Theory Center, Department of Physics, University of
Maryland, College Park, Maryland 20742-4111, USA}

\begin{abstract}
It was proposed that the $id_{x^2-y^2}$ density-wave state (DDW) may
be responsible for the pseudogap behavior in the underdoped
cuprates. Here we show that the admixture of a small $d_{xy}$
component to the DDW state breaks the symmetry between the
counter-propagating orbital currents of the DDW state and, thus,
violates the macroscopic time-reversal symmetry. This
symmetry breaking results in a non-zero polar Kerr effect, which has recently been
observed in the pseudogap phase.
\end{abstract}

\maketitle


\paragraph{Introduction.}


The nature of the pseudogap in the cuprate high-temperature
superconductors is a much-debated and still unresolved issue
\cite{Norman}. Many of the anomalous properties associated with the
pseudogap can be best described as resulting from the order parameters
\cite{Kivelson}, such as those of various density-wave states,
competing with the $d$-wave superconductivity.  If a density-wave
state were to have an energy gap of the $d$-wave symmetry on the Fermi
surface, in conformity with the pseudogap, the natural candidate would
be the $id_{x^2-y^2}$ density wave (DDW) state. Indeed, much of the
phenomenology of the cuprates in the underdoped regime can be unified
\cite{Chakravarty01,Sumanta,Chakravarty04} by making a single
assumption that the ordered DDW state is responsible for the
pseudogap. However, the situation is still controversial, because
there is no convincing direct experimental evidence for a phase
transition to another state in the underdoped regime, besides the
antiferromagnetic order close to half-filling.

The recent observation of quantum oscillations
\cite{Doiron-Leyraud07} in underdoped YBCO in high
magnetic fields has 
renewed interest in the ordered DDW state \cite{Chakravarty07}.
 The special coherence factors of the DDW state 
reduce the hole pockets to appear as Fermi arcs in angle resolved photo emission experiments
\cite{Sumanta-ARPES}. 
An important recent development in the context of the pseudogap is the observation of a non-zero polar
Kerr effect (PKE) in the underdoped YBCO \cite{Kapitulnik07}, which
demonstrates the macroscopic time-reversal-symmetry breaking. The PKE
appears roughly at the same temperature where the pseudogap develops
\cite{Kapitulnik07}, and, near the optimal doping, appears at a
temperature \textit{below} the superconducting transition
temperature. This observation suggests that the time-reversal symmetry
breaking and the pseudogap in the cuprates have the same physical
origin, which is unrelated to the superconductivity. Similar conclusion was reached earlier,
using muon spin relaxation experiments, in Ref.~\cite{Sonier}.  In this Letter,
we examine the PKE of the ordered DDW state, and show that it can give
rise to a non-zero signal only if there is an admixture of a $d_{xy}$
component to the total order parameter amplitude.  There can be
different mechanisms for spontaneous generation of the $d_{xy}$
potential in the mean-field Hamiltonian due to microscopic
interactions between electrons \cite{Nayak} or due to a structural
transition with the appropriate symmetry.  
It is important to stress, however, that the amplitude
of the $d_{xy}$ admixture, as deduced from the PKE signal
\cite{Kapitulnik07}, is tiny relative to the main $id_{x^{2}-y^{2}}$
component. Thus, it may be difficult to uncover it unless the
experiment directly probes broken time-reversal symmetry. Superconducting order parameters
of similar form,
explaining experimental observation of broken time-reversal symmetry, have been discussed in \cite{Luke}.

\paragraph{$id_{x^{2}-y^{2}}$ and $(d+id)$ density-wave states.}

\begin{figure}[t]
\includegraphics[width=0.8\linewidth]{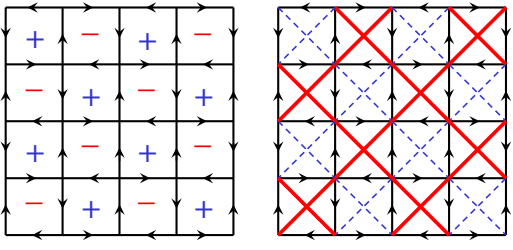}
\caption{Left: The real-space picture of the $id_{x^2-y^2}$ density-wave
state. Spontaneous currents are shown by the arrows. Clockwise and
counterclockwise circulations are indicated by the $-$ and the $+$ signs,
respectively. Right: The same for the $d_{xy}+id_{x^2-y^2}$ density-wave
state. Now the neighboring plaquettes have different amplitudes of the
next-nearest-neighbor tunneling, indicated by the solid and dashed diagonal
lines.}
\label{Fig:d+id}
\end{figure}

The commensurate DDW state \cite{Nayak} is described by the following
order parameter, which is a particle-hole singlet in spin space,
\begin{equation}
\left\langle \hat c_{\bm k+\bm Q,\alpha}^{\dagger} \hat c_{\bm k,\beta}
\right\rangle \propto iW_{\bm k}\,\delta_{\alpha\beta}, \; W_{\bm k}=\frac{
W_0}{2}(\cos k_x-\cos k_y).  \label{Order-Parameter}
\end{equation}
Here $\hat c^{\dagger}$ and $\hat c$ are the electron creation and
annihilation operators on the square lattice of the copper atoms, $\bm
k=(k_x,k_y)$ is the two-dimensional momentum, $\bm Q=(\pi,\pi)$ is the
wave vector of the density wave, and $\alpha$ and $\beta$ are the spin
indices.  Since the order parameter involves $\delta_{\alpha\beta}$,
we do all of our calculations ignoring the spins and just multiply the
final results by two.  We also set the lattice constant $a$, the
electron charge $e$, and the Planck constant $\hbar$ to unity
($e,\hbar,a=1$) at the intermediate stages of the calculations and
restore the full units only in the final results. In
Eq.~(\ref{Order-Parameter}), $iW_{\bm k}$ is the DDW order parameter
with the $id_{x^2-y^2}$ symmetry in the momentum space. For $\bm
Q=(\pi,\pi)$, it is purely imaginary \cite{Nayak} and gives rise to
spontaneous currents along the bonds of the square lattice, as shown
in the left panel of Fig.\ \ref{Fig:d+id}. Although the presence of
the spontaneous currents in the DDW state violates the microscopic
time-reversal symmetry, there is no macroscopic violation. This
happens because the DDW state preserves the combined symmetry of time
reversal and translation by one lattice period. As a result, the
staggered magnetic flux produced by the currents averages to zero on
the macroscopic scale. So, one has to go further in order to explain
the macroscopic time-reversal symmetry breaking revealed by the PKE
measurements \cite{Kapitulnik07}.

We propose to consider a generalization to the $d_{xy}+id_{x^{2}-y^{2}}$
density wave, where the order parameter is a combination of two density
waves with different angular patterns:
\begin{eqnarray}
&& \left\langle \hat c_{\bm k+\bm Q,\alpha}^{\dagger} \hat c_{\bm k,\beta}
\right\rangle \propto [iW_{\bm k}-\Delta_{\bm k}]\,\delta_{\alpha\beta},
\label{Order-Parameter2} \\
&& \Delta_{\bm k}=\Delta_0\sin k_x\sin k_y.
\end{eqnarray}
The real-space structure of this order parameter \cite{Nayak} is shown
in the right panel of Fig.~\ref{Fig:d+id}. The solid and dashed
diagonal lines represent the staggered modulation of the
next-nearest-neighbor electron tunneling due to the $d_{xy}$ component
of the density wave. This staggered modulation breaks the symmetry
between the plaquettes with positive and negative circulations, so the
macroscopic time-reversal symmetry is violated.

The Hartree-Fock Hamiltonian describing the mean-field $(d+id)$-DW is
\begin{eqnarray}
&\hat{H}=\left(
\begin{array}{cc}
\varepsilon _{\bm k}-\mu & iW_{\bm k}-\Delta _{\bm k} \\
-iW_{\bm k}-\Delta _{\bm k} & \varepsilon _{\bm k+\bm Q}-\mu%
\end{array}
\right) ,&  \label{Hamiltonian} \\
&\varepsilon _{\bm k}=-2t(\cos k_{x}+\cos k_{y})+4t^{\prime }\cos k_{x}\cos
k_{y},&
\end{eqnarray}
where $\varepsilon _{\bm k}$ is the band dispersion of the electrons,
and $\mu$ is the chemical potential. The Hamiltonian
(\ref{Hamiltonian}) operates on the two-component spinor
$\hat{\Psi}_{\bm k}=(\hat{c}_{\bm k}, \hat{c}_{\bm k+\bm Q})$ and can
be expanded over the Pauli matrices $\hat{ \bm \tau }$ and the unity
matrix $\hat{I}$, $\hat{H}=w_{0}(\bm k)\hat{I}+\bm w(\bm k)\cdot \hat{\bm\tau },\quad w_{0}=
\frac{\varepsilon _{\bm k}+\varepsilon _{\bm k+\bm Q}}{2}-\mu$, $w_{1}=-\Delta _{\bm k},\quad w_{2}=-W_{\bm k},\quad w_{3}=\frac{
\varepsilon _{\bm k}-\varepsilon _{\bm k+\bm Q}}{2}$.

\begin{figure}[t!]
\includegraphics[width=0.4\linewidth]{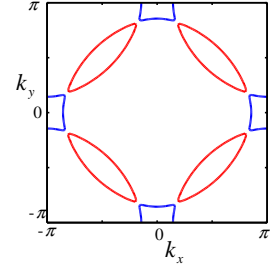}
\caption{The Fermi surface of the $d_{xy}+id_{x^{2}-y^{2}}$
density-wave state, consisting of the electron and hole pockets. The
hole pockets are located near $(\protect\pi /2,\pm \protect\pi /2)$,
and the electron pocket near ($\protect\pi ,0$) in the reduced
Brillouin zone. The band-structure parameters are given at the end of
the paper, and $W_0=20$ meV is taken for this plot. }
\label{Fig:pocket}
\end{figure}

The spectrum of the Hamiltonian
consists of two branches
with the eigenenergies $E_{\pm }(\bm k)=w_{0}(\bm k) \pm w(\bm k)$,
where $w(\bm k)=|\bm w(\bm k)|$. For a generic set of parameters,
corresponding to a non-zero hole doping in the underdoped regime, the
reconstructed Fermi surface consists of two hole pockets near the
$(\pi /2,\pm \pi /2)$ points and one electron pocket near the $(\pi
,0)$ point in the reduced Brillouin zone, as shown in
Fig.~\ref{Fig:pocket}. The Hamiltonian (\ref{Hamiltonian}) with
$t^{\prime }=0$ and $\mu =0$ was studied in Ref.~\cite{Yakovenko90},
where it was found that this system exhibits the intrinsic quantum
Hall effect without an external magnetic field. This confirms that the
time-reversal symmetry is spontaneously broken in the $(d+id)$
density-wave state.

\paragraph{Polar Kerr effect in the $d+id$ density-wave state.}

The polar Kerr angle $\theta _{K}$ can be expressed in terms of the
imaginary part of the ac Hall conductivity $\sigma _{xy}^{\prime
\prime }(\Omega )$ at a frequency $\Omega $ \cite{White},
\begin{equation}
\theta _{K}=\frac{4\pi }{n(n^{2}-1)\Omega d}\,\sigma _{xy}^{\prime \prime
}(\Omega ),  \label{Kerr-Angle}
\end{equation}
where $\sigma _{xy}$ is the Hall conductivity per 2D layer, which is
related to the 3D conductivity by the interlayer distance $d$, and $n$
is the refractive index of the system. The Kerr angle was calculated
for a $p_x+ip_y $ chiral superconductor \cite{Yakovenko07}, motivated
by a non-zero PKE in $\mathrm{Sr_2RuO_4}$ \cite{Kapitulnik06}. Eq.\
(\ref{Kerr-Angle}) is valid when the imaginary part of $n$ is small
relative to the real part.  Otherwise, there is also an additional
contribution \cite{White} from the real part $\sigma _{xy}^{\prime
}(\Omega )$. For simplicity, we focus on $ \sigma_{xy}^{\prime \prime
}(\Omega )$, but $\sigma _{xy}^{\prime }(\Omega ) $ can be also easily
obtained from our calculations.

Introducing the Lagrangian density, $\mathcal{L}=i\omega _{n}
\hat{I}-\hat{H}(\bm k)$, where $\omega _{n}$ is a fermionic Matsubara
frequency, we obtain the electron Green's function as a function of
$\vec{k}=(i\omega _{n},k_{x},k_{y})$
\begin{eqnarray}
&&G_{0}(\vec{k})=\mathcal{L}^{-1}=\frac{[i\omega _{n}-w_{0}(\bm k)]\,\hat{I}+%
\bm w(\bm k)\cdot \hat{\bm\tau }}{g(\omega _{n},\bm k)},
\label{Green-Function} \\
&&g(\vec{k})=[i\omega _{n}-w_{0}(\bm k)]^{2}-|\bm w(\bm k)|^{2}.  \label{g}
\end{eqnarray}
In order to calculate the Hall conductivity, we consider the
electromagnetic potential $\vec{A}=(A_{0},A_{x},A_{y})$, where $A_{0}$
is the scalar potential, and $\bm A=(A_{x},A_{y})$ is the vector
potential. It is introduced into the Lagrangian density $\mathcal{L}$
by the standard substitution $\bm k\rightarrow \bm k-\bm A$ and
$i\omega _{n}\rightarrow i\omega _{n}-A_{0}$, where $c$ is absorbed
into $\bm A$. Expanding $\mathcal{L}$ to the first order in $\vec{A}$,
we find the following addition to the Lagrangian density
\begin{equation}
\Gamma (\vec{q},\vec{k})=-A_{0}(\vec{q})\,\hat{I}+\bm A(\vec{q})\cdot \bm %
\nabla _{\bm k}\; [\bm w(\bm k)\cdot \hat{\bm\tau }].  \label{Gamma}
\end{equation}
Here the electromagnetic potential $\vec{A}\left( {\vec{q}}\right) $
is assumed to be a function of the Fourier variables $\vec{q}=(i\Omega
_{m},q_{x},q_{y})$, where $\Omega _{m}$ is a bosonic Matsubara
frequency.  The Lagrangian (\ref{Gamma}) operates between the
fermionic spinors $\hat{\Psi}(\vec{k})$ and $\hat{\Psi}^{\dagger
}(\vec{k}+\vec{q})$. Calculating the one-loop Feynman diagrams, we
find the following effective action to the second order in $\Gamma $,
\begin{equation}
S=\frac{1}{2}\text{Tr}\sum_{\vec{k},\vec{q}}\Gamma (\vec{q},\vec{k} )G_{0}(
\vec{k}) \Gamma (-\vec{q},\vec{k}+\vec q )G_{0}( \vec{k}+\vec{q}) .
\label{Effective-Action}
\end{equation}
Substituting Eqs.~(\ref{Green-Function}), (\ref{g}), and (\ref{Gamma})
into Eq.~(\ref{Effective-Action}), we write the effective action in
the form
\begin{equation}
S=\sum_{\vec{k},\vec{q}}\frac{\text{Tr}\;\;C_1}{C_2}.
\label{Effective-Action2}
\end{equation}
where $C_1$ is the product of the numerators from
Eq.~(\ref{Green-Function}) and Eq.~(\ref{Gamma}), and
$C_2=g(\vec{k}+\vec q)\,g(\vec{k})$.

Calculating the trace over the $\hat{\tau}$ matrices in the numerator
of Eq.~(\ref{Effective-Action2}), we look for the Chern-Simons-like
terms, which are responsible for the Hall effect. Indeed, we find two
such contributions to the effective action, $S_{H}=S_{1}+S_{2}$,
\begin{eqnarray}
S_{1} &=&i\sum\limits_{\vec{q}}\sigma _{xy}A_{0}(\vec{q})[q_{x}A_{y}(-\vec{q}
)-q_{y}A_{x}(-\vec{q})],  \label{S1} \\
S_{2} &=&\sum\limits_{\vec{q}}\frac{\Omega _{m}}{2}\sigma _{xy}[ A_{x}(\vec{q%
})A_{y}(-\vec{q})-A_{y}(\vec{q})A_{x}(-\vec{q})].  \label{S2}
\end{eqnarray}
By taking the trace in Eq.~(\ref{Effective-Action2}), we find the
following expression for the coefficient $\sigma _{xy}$ in
Eqs.~(\ref{S1}) and (\ref{S2})
\begin{eqnarray}
&& \sigma _{xy}(\Omega _{m})=\sum_{\vec k}\frac{B(\bm k)} {g(\omega_n,\bm %
k)\,g(\omega_n+\Omega_m,\bm k)},  \label{Hall-Conductivity} \\
&& B(\bm k)=-2\bm w\cdot \left[ \frac{\partial \bm w}{\partial k_{x}}\times
\frac{\partial \bm w}{\partial k_{y}}\right] ,  \label{B-w}
\end{eqnarray}
where the vector function $\bm w(\bm k)$ is taken from the Hamiltonian.
 Indeed, the numerator in Eq.~(\ref{Effective-Action2}) is
a product of four terms, each containing the $\hat\tau$ matrices. The
Chern-Simons action (\ref{S1}) and (\ref{S2}) is obtained when these
terms contribute different $\hat\tau$ matrices and the unity matrix to
the trace. Using the formula
$\text{Tr}[\hat\tau_l\hat\tau_m\hat\tau_n]=2i\epsilon_{lmn}$, where
$\epsilon_{lmn}$ is the antisymmetric tensor, we arrive at
Eq.~(\ref{B-w}).  We also expanded the numerator of
Eq.~(\ref{Effective-Action2}) to the first order in $\bm q$ in order
to obtain Eq.\ (\ref{S1}) and then took the limit $\bm q\rightarrow
0$, which is appropriate for the electromagnetic response of
non-relativistic electrons. Substituting the expression for $\bm w(\bm
k)$ from the Hamiltonian into the general formula
(\ref{B-w}), we find $B(\bm k)$ for the $d+id$ density wave, $B(\bm
k)=4t\Delta _{0}W_{0}(\sin ^{2}k_{y}+\cos ^{2}k_{y}\sin ^{2}k_{x})$.
Notice that $B(\bm k)$ is proportional to the product of $\Delta _{0}$
and $W_{0}$, so the intrinsic Hall conductivity is non-zero only when
both components of the $d+id$ density wave are present. The function
$B(\bm k)$ in Eqs.\ (\ref{B-w}) and above is the Berry curvature for
the Hamiltonian \cite{Berry03}.

The coefficient $\sigma _{xy}$ in Eqs.~(\ref{S1}) and (\ref{S2}) is
the Hall conductivity. Indeed, taking the variation
$j_{x}(\vec{q})=\delta S_{H}/\delta A_{x}(-\vec{q})$, we find the Hall
relation at a finite frequency
\begin{equation}
j_{x}(\Omega_m)=\sigma _{xy}(\Omega_m)\,E_{y}(\Omega_m),  \label{Hall}
\end{equation}
where $\bm E=-i\bm q A_{0}-\Omega _{m}\bm A$ is the electric
field. Now we perform summation over the fermionic Matsubara frequency
$\omega_{n}$ in Eq.~(\ref{Hall-Conductivity}) and analytic
continuation to the real bosonic frequency $i\Omega _{m}\rightarrow
\Omega +i\delta $, where $\delta$ is infinitesimal. The result is
\begin{equation}
\sigma _{xy}(\Omega )=\int\limits_{\mathrm{RBZ}}\frac{d^2k}{(2\pi)^2} \frac{%
B(\bm k)\;\{N_F[E_+(\bm k)]-N_F[E_-(\bm k)]\}} {w(\bm k)\,[\Omega+i\delta
-2w(\bm k)]\,[\Omega +i\delta+2w(\bm k)]} ,  \label{s_xy}
\end{equation}
where $N_F(E)$ is the Fermi occupation function for a temperature $T$,
and $2w(\bm k)=E_+(\bm k)-E_-(\bm k)$ is the energy gap between the
bands.  The integral over $\bm k$ is taken over the RBZ.

Let us now explore different limiting cases for
Eq.~(\ref{s_xy}). First we consider the dc limit $\Omega =0$ at $T=0$
in the case when there are no pockets, i.e.\ when $E_+(\bm k)>0$ and
$E_-(\bm k)<0$ for all $\bm k$. This situation takes place, e.g.,\
when $t^{\prime }=0$ and $\mu =0$. In this case, using
Eq.~(\ref{B-w}), we can write Eq.~(\ref{s_xy}) as
\begin{equation}
\sigma _{xy}(0)=-\frac{1}{2\pi}\int\limits_{\mathrm{RBZ}}\frac{d^2k}{4\pi}\; %
\bm n\cdot \left[ \frac{\partial \bm n }{\partial k_{x}}\times \frac{%
\partial \bm n}{\partial k_{y}}\right] ,  \label{s_0}
\end{equation}
where $\bm n(\bm k)=\bm w(\bm k)/w(\bm k)$ is a unit vector. The
integral in Eq.~(\ref{s_0}) is an integer topological invariant, which
represents the degree of mapping from the vector $\bm k$ to the unit
vector $\bm n$. Using Eqs.~(\ref{B-w}) and (\ref{s_0}), we indeed find
the integer quantum Hall effect per one spin component
$\sigma _{xy}(0)=\mathrm{sign}(tW_{0}\Delta _{0})\,\frac{e^{2}}{h}$,
where we restored the dimensional units. The Hall conductivity
does not depend on the magnitude of the
parameters $t$, $W_0$, and $\Delta_0$, but depends on the sign of
their product.

When the pockets are present, the Fermi occupation functions in
Eq.~(\ref{s_xy}) exclude the pockets area from the integration over
$\bm k$, which we represent symbolically as $\overline{\mathrm{RBZ}}$
\begin{equation}
\sigma _{xy}(0)=\frac{e^{2}}{\hbar}\int\limits_{\overline{\mathrm{RBZ}}}
\frac{d^2k}{(2\pi)^2}\frac{B(\bm k)}{4w^3(\bm k)}.  \label{pockets}
\end{equation}
In this case, the Hall conductivity is no longer an integer. At a
finite temperature $T$, the Fermi occupation functions in
Eq.~(\ref{s_xy}) further reduce the Hall conductivity.

Now we turn to the calculation of the imaginary part of the ac Hall
conductivity $\sigma _{xy}^{\prime \prime }(\Omega )$. The imaginary part
comes from the pole in Eq.\ (\ref{s_xy}) at $\Omega =2w(\bm
k)$. This pole represents vertical interband transitions induced by the
electromagnetic wave. Taking the imaginary part of (\ref{s_xy}), we find
\begin{equation}
\sigma _{xy}^{\prime \prime }(\Omega )=\frac{1}{4\pi \Omega ^{2}}
\int\limits_{\overline{\mathrm{RBZ}}} d^2k\;B(\bm k)\; \delta [\Omega -2w(%
\bm k)].  \label{s''}
\end{equation}
The integral over $\bm k$ in Eq.\ (\ref{s''}) is taken along the line
in the Brillouin zone where the condition $\Omega =2w(\bm k)$ is
satisfied.
Eq.\ (\ref{s''}) is valid both for $T=0$ and
$T\neq0$, as long as the condition $\Omega \gg T$ is
satisfied.

\begin{figure}[t]
\includegraphics[width=0.7\linewidth]{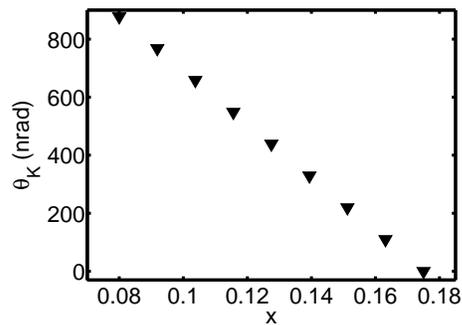}
\caption{The calculated Kerr angle as a function of the hole
concentration $x$. We use the linear dependence
$W_{0}(x)=W_{0}(1-x/x_{c})$ to model roughly the pseudogap variation
with $x$, where $x_{c}=0.175$ and $W_{0}=30$ meV. Although the
admixture $\Delta_{0}=0.001W_{0}/2$ of the $d_{xy}$ component is
small, it leads to a non-zero Kerr angle of the same order of
magnitude as in the experiments \protect\cite{Kapitulnik07}.}
\label{Fig:hallangle}
\end{figure}

Now we make a crude comparison of the theoretical result for the Kerr
angle, given by Eqs.~(\ref{Kerr-Angle}) and (\ref{s''}), with the
experimental measurements \cite{Kapitulnik07}. We use the following
representative band structure parameters: $t=300$ meV, $t^{\prime
}=90$ meV, and $\mu =-260$ meV. The refractive index is $n\approx
1.69$ \cite{Kezuka}, and the interlayer distance is $d=1.17$
nm. Assuming that the pseudogap originates from the ordered DDW state,
we use the formula $W_{0}(x)=30(1-x/x_{c})$ meV with $x_{c}=0.175$, to
model roughly how $W_{0}$ varies with the hole concentration $x$
\cite{Loram}. The optical measurements were performed at the frequency
$\hbar \Omega =0.8$ eV. Taking $\Delta _{0}=\eta W_{0}/2$, with $\eta
=0.1\%$, we plot the value of the Kerr angle $\theta _{K}$ vs.\ the
hole doping $x$ in Fig.\ \ref{Fig:hallangle}, where we have multiplied
the results by 2 to account for the contributions from two spin
components.  The magnitude $\theta _{K}$ roughly corresponds to the
experimental values of the order of $100$ nrad$\ $to $1$ $\mu $rad for
different dopings $x$ \cite{Kapitulnik07}.

The intrinsic dc Hall effect predicted by Eq.~(\ref{pockets}) was not
observed in the underdoped cuprates experimentally. We believe this is
because of the macroscopic domains with opposite chiralities present
in a sample. Their contributions to the dc Hall effect would cancel
out. On the other hand, the optical measurements \cite{Kapitulnik07}
are performed with a small laser beam within a single domain, so they
give a non-zero effect.

Within our theoretical model, the PKE is possible only when both the
$d_{xy}$ and $id_{x^2-y^2}$ order parameters are present. Their
transition temperatures are generally different, so the onset of the
PKE corresponds to the lower of the two transition temperatures.  On
the other hand, the higher transition temperature may be related to
the ``training'' or ``memory'' effect for the sign of the PKE
\cite{Kapitulnik07}.  The transition temperatures of the $d_{xy}$ and
$id_{x^2-y^2}$ density waves should be taken as phenomenological
parameters, because we cannot reliably calculate them from the
microscopic first principles.

In summary, we have developed a minimal theoretical model which can
explain the \textit{macroscopic} time-reversal-symmetry breaking
leading to a non-zero polar Kerr effect \cite{Kapitulnik07} in the
underdoped cuprates. By detailed calculations, we show that admixing a
very small component of the $d_{xy}$ order parameter into the
postulated ordered DDW state picture of the pseudogap produces a PKE
comparable in magnitude to that observed in Ref.~\cite{Kapitulnik07}.
Since the necessary $d_{xy}$ admixture is very small, only of the
order of 0.1\%, all other aspects of the YBCO phenomenology remain
unaffected by our theory.

We thank Aharon Kapitulnik for sharing with us his unpublished data
(\cite{Kapitulnik07}), and for discussion. We thank Roman Lutchyn and
S.~Chakravarty for interesting discussions. This work is supported by
ARO-DARPA and ARO-DTO.




\begin{thebibliography}{99}


\bibitem{Norman} M. R. Norman \textit{et al.}, Adv. Phys. \textbf{54},  715
(2005).

\bibitem{Kivelson} S. A. Kivelson \textit{et al.},  Rev. Mod. Phys. \textbf{%
75}, 1201 (2003).

\bibitem{Chakravarty01} S. Chakravarty \textit{et al.}, Phys. Rev. B
\textbf{63}, 094503 (2001).

\bibitem{Sumanta} S. Tewari \textit{et al.}, Phys. Rev. B \textbf{70},
014514 (2004).

\bibitem{Chakravarty04} S. Chakravarty \textit{et al.}, Nature  \textbf{428}%
, 53 (2004).

\bibitem{Doiron-Leyraud07} N. Doiron-Leyraud \textit{et al.}, Nature \textbf{%
447}, 565 (2007); A. F. Bangura \textit{et al.}, Phys. Rev. Lett. \textbf{%
100}, 047004 (2008).


\bibitem{Chakravarty07} S. Chakravarty and H.-Y. Kee, arXiv:0710.0608
(2007).

\bibitem{Sumanta-ARPES} S. Chakravarty, C. Nayak, and S. Tewari, Phys. Rev B 68,
100504 (2003).






\bibitem{Kapitulnik07} J. Xia \textit{et al.}, arXiv:0711.2494.

\bibitem{Sonier} J. E. Sonier \textit{et al.}, Science \textbf{292}, 1692 (2001).

\bibitem{Nayak} C. Nayak, Phys. Rev. B \textbf{62}, 4880 (2000).


\bibitem{Luke} G. M. Luke \textit{et al.}, Nature \textbf{394}, 558 (1998);
Y. Aoki \textit{et al.}, Phys. Rev. Lett. \textbf{91}, 067003 (2003);
Y. Kasahara \textit{et al.}, Phys. Rev. Lett. \textbf{99}, 116402 (2007).


\bibitem{Yakovenko90} V. M. Yakovenko, Phys. Rev. Lett. \textbf{65},  251
(1990).

\bibitem{White} R. M. White and T. H. Geballe, \textit{Long Range Order in
Solids} (Academic Press, New York, 1979), pp. 317, 321.

\bibitem{Yakovenko07} V. M. Yakovenko, Phys. Rev. Lett. \textbf{98},  087003
(2007).

\bibitem{Kapitulnik06} J. Xia \textit{et al.},  Phys. Rev. Lett. \textbf{97}%
, 167002 (2006).

\bibitem{Berry03} M. V. Berry, Proc. R. Soc. London A \textbf{392}, 92
(2003).

\bibitem{Kezuka} H. Kezuka \textit{et al.}, Physica C \textbf{185},  999
(1991).

\bibitem{Loram} J. W. Loram \textit{et al.}, J. Phys. Chem. Solids \textbf{62%
}, 59 (2001).
\end{thebibliography}
\end{document}